\documentclass[conference]{IEEEtran}

\IEEEoverridecommandlockouts

\usepackage{cite}
\usepackage{amsmath,amsfonts,amssymb,amsthm}
\usepackage{algorithm}
\usepackage{algorithmic}
\usepackage{graphicx}
\usepackage{textcomp}
\usepackage{xcolor}
\usepackage{microtype}
\usepackage{balance}
\usepackage[caption=false,font=footnotesize]{subfig}
\usepackage{tikz}
\usetikzlibrary{shapes.geometric, arrows}
\usetikzlibrary{ arrows.meta, calc, positioning, quotes,}
\usetikzlibrary{arrows.meta,positioning}
\usepackage{adjustbox}
\usepackage{todonotes}
\usepackage{soul}
\usepackage{booktabs}

\tikzstyle{startstop} = [rectangle, rounded corners, minimum width=0.5cm, minimum height=1cm,text centered, draw=black, fill=red!30]
\tikzstyle{io} = [trapezium, trapezium left angle=70, trapezium right angle=110, minimum width=0.5cm, minimum height=1cm, text centered, draw=black, fill=blue!30]
\tikzstyle{process} = [rectangle, minimum width=0.5cm, minimum height=1cm, text centered, text width=3cm, draw=black, fill=orange!30]
\tikzstyle{decision} = [diamond, minimum width=0.5cm, minimum height=1cm, text centered, draw=black, fill=green!30]
\tikzstyle{arrow} = [very thick,->,>=latex]
\tikzstyle{invisible} =[rectangle, node distance=3cm, text=white]
\definecolor{bl1}{HTML}{F7FCF0}
\definecolor{bl2}{HTML}{E0F3DB}
\definecolor{bl3}{HTML}{CCEBC5}
\definecolor{bl4}{HTML}{A8DDB5}
\definecolor{bl5}{HTML}{7BCCC4}

\allowdisplaybreaks 

\makeatletter
\newcommand\fs@betterruled{%
	\def\@fs@cfont{\bfseries}\let\@fs@capt\floatc@ruled
	\def\@fs@pre{\vspace*{5pt}\hrule height.8pt depth0pt \kern2pt}%
	\def\@fs@post{\kern2pt\hrule\relax}%
	\def\@fs@mid{\kern2pt\hrule\kern2pt}%
	\let\@fs@iftopcapt\iftrue}
\floatstyle{betterruled}
\restylefloat{algorithm}
\makeatother

\usepackage{amssymb}
\usepackage{amsfonts}
\usepackage{mathrsfs}
\usepackage{xspace}
\usepackage{bm}
\usepackage{upgreek}

\newcommand{\safemath}[2]{\newcommand{#1}{\ensuremath{#2}\xspace}}

\safemath{\bma}{\mathbf{a}}
\safemath{\bmb}{\mathbf{b}}
\safemath{\bmc}{\mathbf{c}}
\safemath{\bmd}{\mathbf{d}}
\safemath{\bme}{\mathbf{e}}
\safemath{\bmf}{\mathbf{f}}
\safemath{\bmg}{\mathbf{g}}
\safemath{\bmh}{\mathbf{h}}
\safemath{\bmi}{\mathbf{i}}
\safemath{\bmj}{\mathbf{j}}
\safemath{\bmk}{\mathbf{k}}
\safemath{\bml}{\mathbf{l}}
\safemath{\bmm}{\mathbf{m}}
\safemath{\bmn}{\mathbf{n}}
\safemath{\bmo}{\mathbf{o}}
\safemath{\bmp}{\mathbf{p}}
\safemath{\bmq}{\mathbf{q}}
\safemath{\bmr}{\mathbf{r}}
\safemath{\bms}{\mathbf{s}}
\safemath{\bmt}{\mathbf{t}}
\safemath{\bmu}{\mathbf{u}}
\safemath{\bmv}{\mathbf{v}}
\safemath{\bmw}{\mathbf{w}}
\safemath{\bmx}{\mathbf{x}}
\safemath{\bmy}{\mathbf{y}}
\safemath{\bmz}{\mathbf{z}}
\safemath{\bmzero}{\mathbf{0}}
\safemath{\bmone}{\mathbf{1}}

\bmdefine{\biad}{a}
\bmdefine{\bibd}{b}
\bmdefine{\bicd}{c}
\bmdefine{\bidd}{d}
\bmdefine{\bied}{e}
\bmdefine{\bifd}{f}
\bmdefine{\bigd}{g}
\bmdefine{\bihd}{h}
\bmdefine{\biid}{i}
\bmdefine{\bijd}{j}
\bmdefine{\bikd}{k}
\bmdefine{\bild}{l}
\bmdefine{\bimd}{m}
\bmdefine{\bind}{n}
\bmdefine{\biod}{o}
\bmdefine{\bipd}{p}
\bmdefine{\biqd}{q}
\bmdefine{\bird}{r}
\bmdefine{\bisd}{s}
\bmdefine{\bitd}{t}
\bmdefine{\biud}{u}
\bmdefine{\bivd}{v}
\bmdefine{\biwd}{w}
\bmdefine{\bixd}{x}
\bmdefine{\biyd}{y}
\bmdefine{\bizd}{z}

\bmdefine{\bixid}{\xi}
\bmdefine{\bilambdad}{\lambda}
\bmdefine{\bimud}{\mu}
\bmdefine{\bithetad}{\theta}
\bmdefine{\biphid}{\phi}
\bmdefine{\bideltad}{\delta}

\safemath{\bmia}{\biad}
\safemath{\bmib}{\bibd}
\safemath{\bmic}{\bicd}
\safemath{\bmid}{\bidd}
\safemath{\bmie}{\bied}
\safemath{\bmif}{\bifd}
\safemath{\bmig}{\bigd}
\safemath{\bmih}{\bihd}
\safemath{\bmii}{\biid}
\safemath{\bmij}{\bijd}
\safemath{\bmik}{\bikd}
\safemath{\bmil}{\bild}
\safemath{\bmim}{\bimd}
\safemath{\bmin}{\bind}
\safemath{\bmio}{\biod}
\safemath{\bmip}{\bipd}
\safemath{\bmiq}{\biqd}
\safemath{\bmir}{\bird}
\safemath{\bmis}{\bisd}
\safemath{\bmit}{\bitd}
\safemath{\bmiu}{\biud}
\safemath{\bmiv}{\bivd}
\safemath{\bmiw}{\biwd}
\safemath{\bmix}{\bixd}
\safemath{\bmiy}{\biyd}
\safemath{\bmiz}{\bizd}

\safemath{\bmxi}{\bixid}
\safemath{\bmlambda}{\bilambdad}
\safemath{\bmmu}{\bimud}
\safemath{\bmtheta}{\bithetad}
\safemath{\bmphi}{\biphid}
\safemath{\bmdelta}{\bideltad}

\safemath{\bA}{\mathbf{A}}
\safemath{\bB}{\mathbf{B}}
\safemath{\bC}{\mathbf{C}}
\safemath{\bD}{\mathbf{D}}
\safemath{\bE}{\mathbf{E}}
\safemath{\bF}{\mathbf{F}}
\safemath{\bG}{\mathbf{G}}
\safemath{\bH}{\mathbf{H}}
\safemath{\bI}{\mathbf{I}}
\safemath{\bJ}{\mathbf{J}}
\safemath{\bK}{\mathbf{K}}
\safemath{\bL}{\mathbf{L}}
\safemath{\bM}{\mathbf{M}}
\safemath{\bN}{\mathbf{N}}
\safemath{\bO}{\mathbf{O}}
\safemath{\bP}{\mathbf{P}}
\safemath{\bQ}{\mathbf{Q}}
\safemath{\bR}{\mathbf{R}}
\safemath{\bS}{\mathbf{S}}
\safemath{\bT}{\mathbf{T}}
\safemath{\bU}{\mathbf{U}}
\safemath{\bV}{\mathbf{V}}
\safemath{\bW}{\mathbf{W}}
\safemath{\bX}{\mathbf{X}}
\safemath{\bY}{\mathbf{Y}}
\safemath{\bZ}{\mathbf{Z}}

\safemath{\bZero}{\mathbf{0}}
\safemath{\bOne}{\mathbf{1}}
\safemath{\bDelta}{\mathbf{\Delta}}
\safemath{\bLambda}{\mathbf{\UpLambda}}
\safemath{\bPhi}{\mathbf{\Phi}}
\safemath{\bPsi}{\mathbf{\Psi}}
\safemath{\bSigma}{\mathbf{\Upsigma}}
\safemath{\bOmega}{\mathbf{\Upomega}}
\safemath{\bTheta}{\mathbf{\Uptheta}}

\bmdefine{\biAd}{A}
\bmdefine{\biBd}{B}
\bmdefine{\biCd}{C}
\bmdefine{\biDd}{D}
\bmdefine{\biEd}{E}
\bmdefine{\biFd}{F}
\bmdefine{\biGd}{G}
\bmdefine{\biHd}{H}
\bmdefine{\biId}{I}
\bmdefine{\biJd}{J}
\bmdefine{\biKd}{K}
\bmdefine{\biLd}{L}
\bmdefine{\biMd}{M}
\bmdefine{\biOd}{N}
\bmdefine{\biPd}{O}
\bmdefine{\biQd}{P}
\bmdefine{\biRd}{R}
\bmdefine{\biSd}{S}
\bmdefine{\biTd}{T}
\bmdefine{\biUd}{U}
\bmdefine{\biVd}{V}
\bmdefine{\biWd}{W}
\bmdefine{\biXd}{X}
\bmdefine{\biYd}{Y}
\bmdefine{\biZd}{Z}

\bmdefine{\biDelta}{\Delta}
\bmdefine{\biLambda}{\Lambda}
\bmdefine{\biPhi}{\Phi}
\bmdefine{\biSigma}{\Sigma}
\bmdefine{\biOmega}{\Omega}
\bmdefine{\biTheta}{\Theta}

\safemath{\bimA}{\biAd}
\safemath{\bimB}{\biBd}
\safemath{\bimC}{\biCd}
\safemath{\bimD}{\biDd}
\safemath{\bimE}{\biEd}
\safemath{\bimF}{\biFd}
\safemath{\bimG}{\biGd}
\safemath{\bimH}{\biHd}
\safemath{\bimI}{\biId}
\safemath{\bimJ}{\biJd}
\safemath{\bimK}{\biKd}
\safemath{\bimL}{\biLd}
\safemath{\bimM}{\biMd}
\safemath{\bimN}{\biNd}
\safemath{\bimO}{\biOd}
\safemath{\bimP}{\biPd}
\safemath{\bimQ}{\biQd}
\safemath{\bimR}{\biRd}
\safemath{\bimS}{\biSd}
\safemath{\bimT}{\biTd}
\safemath{\bimU}{\biUd}
\safemath{\bimV}{\biVd}
\safemath{\bimW}{\biWd}
\safemath{\bimX}{\biXd}
\safemath{\bimY}{\biYd}
\safemath{\bimZ}{\biZd}

\safemath{\bimDelta}{\biDelta}
\safemath{\bimLambda}{\biLambda}
\safemath{\bimPhi}{\biPhi}
\safemath{\bimSigma}{\biSigma}
\safemath{\bimOmega}{\biOmega}
\safemath{\bimTheta}{\biTheta}

\safemath{\setA}{\mathcal{A}}
\safemath{\setB}{\mathcal{B}}
\safemath{\setC}{\mathcal{C}}
\safemath{\setD}{\mathcal{D}}
\safemath{\setE}{\mathcal{E}}
\safemath{\setF}{\mathcal{F}}
\safemath{\setG}{\mathcal{G}}
\safemath{\setH}{\mathcal{H}}
\safemath{\setI}{\mathcal{I}}
\safemath{\setJ}{\mathcal{J}}
\safemath{\setK}{\mathcal{K}}
\safemath{\setL}{\mathcal{L}}
\safemath{\setM}{\mathcal{M}}
\safemath{\setN}{\mathcal{N}}
\safemath{\setO}{\mathcal{O}}
\safemath{\setP}{\mathcal{P}}
\safemath{\setQ}{\mathcal{Q}}
\safemath{\setR}{\mathcal{R}}
\safemath{\setS}{\mathcal{S}}
\safemath{\setT}{\mathcal{T}}
\safemath{\setU}{\mathcal{U}}
\safemath{\setV}{\mathcal{V}}
\safemath{\setW}{\mathcal{W}}
\safemath{\setX}{\mathcal{X}}
\safemath{\setY}{\mathcal{Y}}
\safemath{\setZ}{\mathcal{Z}}
\safemath{\emptySet}{\varnothing}

\safemath{\colA}{\mathscr{A}}
\safemath{\colB}{\mathscr{B}}
\safemath{\colC}{\mathscr{C}}
\safemath{\colD}{\mathscr{D}}
\safemath{\colE}{\mathscr{E}}
\safemath{\colF}{\mathscr{F}}
\safemath{\colG}{\mathscr{G}}
\safemath{\colH}{\mathscr{H}}
\safemath{\colI}{\mathscr{I}}
\safemath{\colJ}{\mathscr{J}}
\safemath{\colK}{\mathscr{K}}
\safemath{\colL}{\mathscr{L}}
\safemath{\colM}{\mathscr{M}}
\safemath{\colN}{\mathscr{N}}
\safemath{\colO}{\mathscr{O}}
\safemath{\colP}{\mathscr{P}}
\safemath{\colQ}{\mathscr{Q}}
\safemath{\colR}{\mathscr{R}}
\safemath{\colS}{\mathscr{S}}
\safemath{\colT}{\mathscr{T}}
\safemath{\colU}{\mathscr{U}}
\safemath{\colV}{\mathscr{V}}
\safemath{\colW}{\mathscr{W}}
\safemath{\colX}{\mathscr{X}}
\safemath{\colY}{\mathscr{Y}}
\safemath{\colZ}{\mathscr{Z}}

\safemath{\opA}{\mathbb{A}}
\safemath{\opB}{\mathbb{B}}
\safemath{\opC}{\mathbb{C}}
\safemath{\opD}{\mathbb{D}}
\safemath{\opE}{\mathbb{E}}
\safemath{\opF}{\mathbb{F}}
\safemath{\opG}{\mathbb{G}}
\safemath{\opH}{\mathbb{H}}
\safemath{\opI}{\mathbb{I}}
\safemath{\opJ}{\mathbb{J}}
\safemath{\opK}{\mathbb{K}}
\safemath{\opL}{\mathbb{L}}
\safemath{\opM}{\mathbb{M}}
\safemath{\opN}{\mathbb{N}}
\safemath{\opO}{\mathbb{O}}
\safemath{\opP}{\mathbb{P}}
\safemath{\opQ}{\mathbb{Q}}
\safemath{\opR}{\mathbb{R}}
\safemath{\opS}{\mathbb{S}}
\safemath{\opT}{\mathbb{T}}
\safemath{\opU}{\mathbb{U}}
\safemath{\opV}{\mathbb{V}}
\safemath{\opW}{\mathbb{W}}
\safemath{\opX}{\mathbb{X}}
\safemath{\opY}{\mathbb{Y}}
\safemath{\opZ}{\mathbb{Z}}
\safemath{\opZero}{\mathbb{O}}
\safemath{\identityop}{\opI}

\safemath{\veca}{\bma}
\safemath{\vecb}{\bmb}
\safemath{\vecc}{\bmc}
\safemath{\vecd}{\bmd}
\safemath{\vece}{\bme}
\safemath{\vecf}{\bmf}
\safemath{\vecg}{\bmg}
\safemath{\vech}{\bmh}
\safemath{\veci}{\bmi}
\safemath{\vecj}{\bmj}
\safemath{\veck}{\bmk}
\safemath{\vecl}{\bml}
\safemath{\vecm}{\bmm}
\safemath{\vecn}{\bmn}
\safemath{\veco}{\bmo}
\safemath{\vecp}{\bmp}
\safemath{\vecq}{\bmq}
\safemath{\vecr}{\bmr}
\safemath{\vecs}{\bms}
\safemath{\vect}{\bmt}
\safemath{\vecu}{\bmu}
\safemath{\vecv}{\bmv}
\safemath{\vecw}{\bmw}
\safemath{\vecx}{\bmx}
\safemath{\vecy}{\bmy}
\safemath{\vecz}{\bmz}

\safemath{\veczero}{\bmzero}
\safemath{\vecone}{\bmone}
\safemath{\vecxi}{\bmxi}
\safemath{\veclambda}{\bmlambda}
\safemath{\vecmu}{\bmmu}
\safemath{\vectheta}{\bmtheta}
\safemath{\vecphi}{\bmphi}
\safemath{\vecdelta}{\bmdelta}

\safemath{\matA}{\bA}
\safemath{\matB}{\bB}
\safemath{\matC}{\bC}
\safemath{\matD}{\bD}
\safemath{\matE}{\bE}
\safemath{\matF}{\bF}
\safemath{\matG}{\bG}
\safemath{\matH}{\bH}
\safemath{\matI}{\bI}
\safemath{\matJ}{\bJ}
\safemath{\matK}{\bK}
\safemath{\matL}{\bL}
\safemath{\matM}{\bM}
\safemath{\matN}{\bN}
\safemath{\matO}{\bO}
\safemath{\matP}{\bP}
\safemath{\matQ}{\bQ}
\safemath{\matR}{\bR}
\safemath{\matS}{\bS}
\safemath{\matT}{\bT}
\safemath{\matU}{\bU}
\safemath{\matV}{\bV}
\safemath{\matW}{\bW}
\safemath{\matX}{\bX}
\safemath{\matY}{\bY}
\safemath{\matZ}{\bZ}
\safemath{\matzero}{\bmzero}

\safemath{\matDelta}{\bDelta}
\safemath{\matLambda}{\bLambda}
\safemath{\matPhi}{\bPhi}
\safemath{\matSigma}{\bSigma}
\safemath{\matOmega}{\bOmega}
\safemath{\matTheta}{\bTheta}

\safemath{\matidentity}{\matI}
\safemath{\matone}{\matO}

\safemath{\rnda}{A}
\safemath{\rndb}{B}
\safemath{\rndc}{C}
\safemath{\rndd}{D}
\safemath{\rnde}{E}
\safemath{\rndf}{F}
\safemath{\rndg}{G}
\safemath{\rndh}{H}
\safemath{\rndi}{I}
\safemath{\rndj}{J}
\safemath{\rndk}{K}
\safemath{\rndl}{L}
\safemath{\rndm}{M}
\safemath{\rndn}{N}
\safemath{\rndo}{O}
\safemath{\rndp}{P}
\safemath{\rndq}{Q}
\safemath{\rndr}{R}
\safemath{\rnds}{S}
\safemath{\rndt}{T}
\safemath{\rndu}{U}
\safemath{\rndv}{V}
\safemath{\rndw}{W}
\safemath{\rndx}{X}
\safemath{\rndy}{Y}
\safemath{\rndz}{Z}

\safemath{\rveca}{\bimA}
\safemath{\rvecb}{\bimB}
\safemath{\rvecc}{\bimC}
\safemath{\rvecd}{\bimD}
\safemath{\rvece}{\bimE}
\safemath{\rvecf}{\bimF}
\safemath{\rvecg}{\bimG}
\safemath{\rvech}{\bimH}
\safemath{\rveci}{\bimI}
\safemath{\rvecj}{\bimJ}
\safemath{\rveck}{\bimK}
\safemath{\rvecl}{\bimL}
\safemath{\rvecm}{\bimM}
\safemath{\rvecn}{\bimN}
\safemath{\rveco}{\bomO}
\safemath{\rvecp}{\bimP}
\safemath{\rvecq}{\bimQ}
\safemath{\rvecr}{\bimR}
\safemath{\rvecs}{\bimS}
\safemath{\rvect}{\bimT}
\safemath{\rvecu}{\bimU}
\safemath{\rvecv}{\bimV}
\safemath{\rvecw}{\bimW}
\safemath{\rvecx}{\bimX}
\safemath{\rvecy}{\bimY}
\safemath{\rvecz}{\bimZ}

\safemath{\rvecxi}{\bmxi}
\safemath{\rveclambda}{\bmlambda}
\safemath{\rvecmu}{\bmmu}
\safemath{\rvectheta}{\bmtheta}
\safemath{\rvecphi}{\bmphi}

\safemath{\rmatA}{\bimA}
\safemath{\rmatB}{\bimB}
\safemath{\rmatC}{\bimC}
\safemath{\rmatD}{\bimD}
\safemath{\rmatE}{\bimE}
\safemath{\rmatF}{\bimF}
\safemath{\rmatG}{\bimG}
\safemath{\rmatH}{\bimH}
\safemath{\rmatI}{\bimI}
\safemath{\rmatJ}{\bimJ}
\safemath{\rmatK}{\bimK}
\safemath{\rmatL}{\bimL}
\safemath{\rmatM}{\bimM}
\safemath{\rmatN}{\bimN}
\safemath{\rmatO}{\bimO}
\safemath{\rmatP}{\bimP}
\safemath{\rmatQ}{\bimQ}
\safemath{\rmatR}{\bimR}
\safemath{\rmatS}{\bimS}
\safemath{\rmatT}{\bimT}
\safemath{\rmatU}{\bimU}
\safemath{\rmatV}{\bimV}
\safemath{\rmatW}{\bimW}
\safemath{\rmatX}{\bimX}
\safemath{\rmatY}{\bimY}
\safemath{\rmatZ}{\bimZ}

\safemath{\rmatDelta}{\bimDelta}
\safemath{\rmatLambda}{\bimLambda}
\safemath{\rmatPhi}{\bimPhi}
\safemath{\rmatSigma}{\bimSigma}
\safemath{\rmatOmega}{\bimOmega}
\safemath{\rmatTheta}{\bimTheta}

\usepackage{amssymb}
\usepackage{amsfonts}
\usepackage{mathrsfs}
\usepackage{xspace}
\usepackage{bm}
\usepackage{fancyref}
\usepackage{textcomp}

\usepackage{multirow}
\usepackage{stmaryrd}

\newenvironment{textbmatrix}{	\setlength{\arraycolsep}{2.5pt}%
								\big[\begin{matrix}}{\end{matrix}\big]%
								\raisebox{0.08ex}{\vphantom{M}}}

\def\be{\begin{equation}}
\def\ee{\end{equation}}
\def\een{\nonumber \end{equation}}
\def\mat{\begin{bmatrix}}
\def\emat{\end{bmatrix}}
\def\btm{\begin{textbmatrix}}
\def\etm{\end{textbmatrix}}

\def\ba#1\ea{\begin{align}#1\end{align}}
\def\bas#1\eas{\begin{align*}#1\end{align*}}
\def\bs#1\es{\begin{split}#1\end{split}} 
\def\bg#1\eg{\begin{gather}#1\end{gather}}
\def\bml#1\eml{\begin{multline}#1\end{multline}}
\def\bi#1\ei{\begin{itemize}#1\end{itemize}}

\safemath{\dirac}{\delta}					
\safemath{\krond}{\dirac}					

\safemath{\upto}{\uparrow}
\safemath{\downto}{\downarrow}
\safemath{\iu}{j}							
\safemath{\ev}{\lambda}						
\safemath{\hilseqspace}{l^{2}}				
\newcommand{\banachfunspace}[1]{\setL^{#1}}	
\safemath{\hilfunspace}{\banachfunspace{2}}	

\safemath{\SNR}{\text{\sc snr}} 				
\safemath{\No}{N_0}							
\safemath{\Es}{E_s}							
\safemath{\Eb}{E_b}							
\safemath{\EbNo}{\frac{\Eb}{\No}}
\safemath{\EsNo}{\frac{\Es}{\No}}

\DeclareMathOperator{\CHop}{\ensuremath{\opH}} 
\safemath{\tvir}{\rndh_{\CHop}}				
\safemath{\tvtf}{\rndl_{\CHop}}				
\safemath{\spf}{\rnds_{\CHop}}				
\safemath{\bff}{H_{\CHop}}					

\safemath{\ircf}{r_{h}}						
\safemath{\tftvcf}{r_{s}}					
\safemath{\tfcf}{r_{l}}						
\safemath{\bfcf}{r_{H}}						

\safemath{\tcorr}{c_h}						
\safemath{\scf}{c_{s}}						
\safemath{\tfcorr}{c_{l}}					
\safemath{\fcorr}{c_{H}}						

\safemath{\mi}{I}							
\safemath{\capacity}{C}						

\safemath{\normal}{\mathcal{N}}			
\safemath{\jpg}{\mathcal{CN}}			
\safemath{\mchain}{\leftrightarrow}		

\safemath{\dB}{\,\mathrm{dB}}
\safemath{\dBm}{\,\mathrm{dBm}}
\safemath{\Hz}{\,\mathrm{Hz}}
\safemath{\kHz}{\,\mathrm{kHz}}
\safemath{\MHz}{\,\mathrm{MHz}}
\safemath{\GHz}{\,\mathrm{GHz}}
\safemath{\s}{\,\mathrm{s}}
\safemath{\ms}{\,\mathrm{ms}}
\safemath{\mus}{\,\mathrm{\text{\textmu}s}}
\safemath{\ns}{\,\mathrm{ns}}
\safemath{\ps}{\,\mathrm{ps}}
\safemath{\meter}{\,\mathrm{m}}
\safemath{\mm}{\,\mathrm{mm}}
\safemath{\cm}{\,\mathrm{cm}}
\safemath{\W}{\,\mathrm{W}}
\safemath{\mW}{\, \mathrm{mW}}
\safemath{\J}{\,\mathrm{J}}
\safemath{\K}{\,\mathrm{K}}
\safemath{\bit}{\,\mathrm{bit}}
\safemath{\nat}{\,\mathrm{nat}}

\safemath{\define}{\triangleq}			

\safemath{\equivalent}{\sim}
\safemath{\distas}{\sim}					
\safemath{\sdiff}{\Delta}				

\safemath{\reals}{\mathbb{R}}
\safemath{\positivereals}{\reals_{+}}
\safemath{\integers}{\mathbb{Z}}
\safemath{\posint}{\integers_{+}}
\safemath{\naturals}{\mathbb{N}}
\safemath{\posnaturals}{\naturals_{+}}
\safemath{\complexset}{\mathbb{C}}
\safemath{\rationals}{\mathbb{Q}}

\newcommand*{\fancyrefapplabelprefix}{app}		
\newcommand*{\fancyrefthmlabelprefix}{thm}		
\newcommand*{\fancyreflemlabelprefix}{lem}		
\newcommand*{\fancyrefcorlabelprefix}{cor}		
\newcommand*{\fancyrefdeflabelprefix}{def}		
\newcommand*{\fancyrefproplabelprefix}{prop}		
\newcommand*{\fancyrefexmpllabelprefix}{exmpl}
\frefformat{vario}{\fancyrefseclabelprefix}{Section~#1}
\frefformat{vario}{\fancyrefthmlabelprefix}{Theorem~#1}
\frefformat{vario}{\fancyreflemlabelprefix}{Lemma~#1}
\frefformat{vario}{\fancyrefcorlabelprefix}{Corollary~#1}
\frefformat{vario}{\fancyrefdeflabelprefix}{Definition~#1}
\frefformat{vario}{\fancyreffiglabelprefix}{Fig.~#1}
\frefformat{vario}{\fancyrefapplabelprefix}{Appendix~#1}
\frefformat{vario}{\fancyrefeqlabelprefix}{(#1)}
\frefformat{vario}{\fancyrefproplabelprefix}{Property~#1}
\frefformat{vario}{\fancyrefexmpllabelprefix}{Example~#1}

\def\s{{\mathbf s}}

\def\A{{\mathbf A}}

\def\W{{\mathbf W}}

\def\MR{{M_\text{R}}}
\begin{document}
	
\title{Feature Learning for Neural-Network-Based Positioning with Channel State Information}
	
\author{\IEEEauthorblockN{Emre G\"{o}n\"{u}lta\c{s}$^\text{1}$, Sueda Taner$^\text{2}$, Howard Huang$^\text{3}$, and Christoph Studer$^\text{2}$}\\[-0.1cm]
\IEEEauthorblockN{\em $^\text{1}$Ericsson, Austin, TX\\
$^\text{2}$Department of Information Technology and Electrical Engineering, ETH Z\"urich, Switzerland\\
$^\text{3}$Nokia Bell-Labs, Murray Hill, NJ 
}\thanks{
EG was with the School of Electrical and Computer Engineering, Cornell University, Ithaca, NY 14853 USA, and is now with Ericsson, Austin, TX 78704, USA.
ST was with the School of Electrical and Computer Engineering, Cornell University, Ithaca, NY 14853 USA, and is now with Department of Information Technology and Electrical Engineering at ETH Z\"urich, Switzerland.}
\thanks{
The work of EG and CS was supported in part by the US National Science Foundation (NSF) under grants CNS-1717559, CNS-1955997, and ECCS-1824379. The work of ST and CS was supported in part by  ComSenTer, one of six centers in JUMP, a SRC program sponsored by DARPA. The work of CS was also supported by an ETH Research Grant.}\thanks{Contact author: E. G\"{o}n\"{u}lta\c{s}; e-mail: eg566@cornell.edu}\thanks{We thank P.~Huang and B.~Rappaport for discussions on DNN-based positioning. We also thank Prof.\ K.~Petersen for allowing  us to use the VICON positioning system~\cite{vicon} and the lab space at Cornell University.}
}

	\maketitle
	
\begin{abstract}
Recent channel state information (CSI)-based positioning pipelines rely on deep neural networks (DNNs) in order to learn a mapping from estimated CSI to position. Since real-world communication transceivers suffer from hardware impairments, CSI-based positioning systems typically rely on features that are designed by hand. In this paper, we propose a CSI-based positioning pipeline that directly takes raw CSI measurements and learns features using a structured DNN in order to generate probability maps describing the likelihood of the transmitter being at pre-defined grid points. To further improve the positioning accuracy of moving user equipments, we propose to fuse a time-series of learned CSI features or a time-series of probability maps. To demonstrate the efficacy of our methods, we perform experiments with real-world  indoor line-of-sight (LoS) and non-LoS channel measurements. We show that CSI feature learning and time-series fusion can reduce the mean distance error by up to 2.5$\boldsymbol\times$ compared to the state-of-the-art.
\end{abstract}
	
\section{Introduction}

The need for low-cost but accurate positioning systems is driven by recent trends in virtual reality, asset tracking, robotics, and industrial automation~\cite{han2016survey,fallah2013indoorsurvey}.
Existing outdoor positioning solutions mostly rely on global navigation satellite systems (GNSS) that provide meter-level accuracy but require line-of-sight (LoS) satellite connectivity. 
High-precision indoor positioning solutions typically require specialized hardware that uses visible or infra-red light to localize objects with either active IR transmitting markers~\cite{worldviz} or passive reflectors~\cite{vicon}.
Such systems require unobstructed views, are affected by sunlight and reflective surfaces, and are costly. 

\subsection{CSI-Based Positioning with Neural Networks}

Low-cost indoor positioning can be achieved with existing communication infrastructure that utilizes orthogonal frequency division multiplexing (OFDM)~\cite{ofdm}. OFDM receivers must acquire channel state information (CSI) to suppress  inter-symbol interference caused by multi-path propagation~\cite{ofdm}.
Since the measured CSI strongly depends on the environment between the transmitting user equipment (UE) and the receiver (access point or base station), it can be used for positioning purposes~\cite{wu2013indoor,yang2013rssi,larsson2015fingerprinting,wang2015deepfi,wang2017csi,lundpaper,arnold2019novel,ericpaper,huawei2020paper,li2020wireless,emrepaper,foliadis2021csibased}.
Such CSI-based positioning solutions either use geometrical models~\cite{WEN201921survey} or learn a function that maps CSI to position~\cite{lundpaper,ericpaper,chen2017confi,huawei2020paper,emrepaper}. 
The latter approach often relies upon deep neural networks (DNNs), which has shown to enable accurate outdoor and indoor positioning accuracy~\cite{wang2015deepfi,arnold2019novel,li2020wireless,huawei2020paper,emrepaper,foliadis2021csibased}.

DNN-based positioning pipelines that rely on \emph{raw} CSI measurements enable sub-meter-level indoor positioning accuracy~\cite{li2020wireless}, but are unable to compete with approaches that use carefully-engineered CSI features~\cite{emrepaper,foliadis2021csibased}.
The main reason is as follows: Real-world CSI measurements are affected by small-scale fading, antenna orientation, and hardware impairments~\cite{huawei2020paper,emrepaper}, such as synchronization errors, residual timing and carrier frequency offsets, and phase noise, which necessitates features that are resilient to such effects.
As a consequence, learning DNNs from raw CSI measurements would require a prohibitive amount of training data to learn from a sufficiently diverse training set that contains all possible combinations of real-world system and hardware nonidealities. 
Hence, existing systems almost exclusively rely on CSI features that are designed by hand and tailored to the communication standard and hardware equipment to be resilient to such impairments.

State-of-the-art CSI feature extraction pipelines often transform the frequency-domain CSI to the delay domain~\cite{lundpaper,ericpaper,huawei2020paper,emrepaper}.
Computing an autocorrelation instead of using the raw delay-domain information has been shown in~\cite{huawei2020paper} to further improve resilience to hardware impairments. 
Although such features combined with DNNs enable centimeter (cm)-level indoor positioning accuracy~\cite{emrepaper}, they do not exploit the learning capabilities of DNNs. 
Furthermore, most CSI feature extraction pipelines compute a position estimate based on  measurements acquired only during a single time instance, with the exception of the recent papers~\cite{li2020wireless,hoang2020cnnlstm}.
However, both of these papers propose to fuse a time series of CSI measurements without providing any insight on the efficacy of fusing the \emph{outputs} of the positioning DNN.

\subsection{Contributions} 

In this paper, we propose a DNN-based positioning pipeline that takes in a time-series of \emph{raw} CSI measurements in order to generate a probability map, which indicates the likelihood of the transmitting user equipment (UE) to be at a certain grid point. 
This probability map is then used to estimate the UE's position.  
In order to enable CSI feature learning, we build upon the structure of hand-designed feature extraction pipelines~\cite{huawei2020paper,emrepaper} while learning its key parameters.
We furthermore improve the positioning accuracy by (i) CSI feature fusion, which combines a time-series of raw-CSI measurements using a recursive neural network (RNN), and (ii) probability map fusion, which combines a time-series of the generated probability maps using an RNN or probability conflation.
We systematically study the efficacy of our methods using real-world CSI measurements for LoS and non-LoS scenarios, and we show that our methods can improve positioning accuracy by up to $2.5\times$ compared to the state-of-the-art.

\subsection{Notation}

Lowercase and uppercase boldface letters denote column vectors and matrices, respectively.
For a matrix~$\bA$, we denote the transpose by~$\bA^T$, Hermitian transpose by~$\A^H$, entry in the $i$th row and $j$th column by $A_{i,j}$, $i$th column by~$\bm{a}_i$, and real and imaginary parts by $\Re(\bA)$ and $\Im(\bA)$, respectively.
The column-wise vectorization of $\bA$ is denoted by $\mathrm{vec}(\bA)$.   
For a vector~$\bma$, the $k$th entry is~$a_k$ and the $\ell^2$-norm is $\|\bma\|_2$.
The operator $|\cdot|^{\circ2}$ is the element-wise absolute value squared.

\section{System Model}

We start by outlining the operation principle of the proposed CSI-based positioning pipeline. We then describe state-of-the-art CSI-based feature extraction and DNN pipelines. 

\subsection{System Model}
We consider a single-input multiple-output (SIMO) OFDM communication system building on the IEEE 802.11ac standard~\cite{ieee80211ac} with $U$ single-antenna mobile UEs, an access point (AP) with~$\MR$ antennas, and OFDM transmission with $W$ occupied subcarriers.
We assume that the $u$th UE at time instant $t$ is at position $\bmx^{(u_t)}\in\reals^D$, where~$D$ is either two or three. 
The UE transmits a pre-defined pilot sequence, which is used by the AP to estimate the CSI on the occupied OFDM subcarriers. Since the system is assumed to operate in time-division duplexing mode, the other UEs are inactive. 
The estimated $\MR$-dimensional SIMO channel vector associated with UE~$u$ at time~$t$ and subcarrier $w=1,\ldots,W$ is denoted by $\bmh^{(u_t)}_w\in\complexset^{\MR}$. We call the collection $\bH^{(u_t)}=[\bmh_1^{(u_t)},\dots,\bmh_W^{(u_t)}]$ of these channel vectors for all subcarriers the \emph{estimated CSI} of UE~$u$ at time~$t$. 
In what follows, we assume that the estimated CSI is affected by real-world system and hardware impairments, such as noise, synchronization errors, mismatches in carrier frequency and sampling rates, and phase noise. 

\begin{figure}[tp]

\centering
\subfloat[\label{fig:rawcsi}]
{
	\adjustbox{max width=0.28\columnwidth,valign=t}{%
		\begin{tikzpicture}[font=\small,node distance=3cm]
		
		\node (start) [startstop,fill=bl1] {\large{Access point}};

		\node (pro2a) [process, below of =start,yshift=-1cm,fill=magenta,fill=bl3] {\large Neural network $g_{\boldsymbol\theta}(\cdot)$};
		\node (pos) [process, below of =pro2a,yshift=1cm,fill=bl4] {\large Extract position};
		\node [invisible, below of=pos,yshift=1.3cm] (inv) {};
		
		\draw [arrow] (start) --node[anchor=east] {\large$\bH^{(u_t)}$}(pro2a);
		\draw [arrow] (pro2a) --node[anchor=east] {\large$\hat{\bmp}^{(u_t)}$} (pos);
		\draw [arrow] (pos) --node[anchor=east] {\large$\hat\bmx^{(u_t)}$}(inv) ;
		
		\end{tikzpicture}
	}
} \hspace{0.2cm}
\subfloat[\label{fig:featuredesign}]
{
	\adjustbox{max width=0.28\columnwidth,valign=t}{%
		\begin{tikzpicture}[font=\small,node distance=3cm]
		
		\node (start) [startstop,fill=bl1] {\large{Access point}};
		\node (pro1) [process, below of=start,yshift=1cm,fill=bl2] {\large Feature design $f(\cdot)$};
		\node (pro2a) [process, below of =pro1,yshift=1cm,fill=magenta,fill=bl3] {\large Neural network $g_{\boldsymbol\theta}(\cdot)$};
		\node (pos) [process, below of =pro2a,yshift=1cm,fill=bl4] {\large Extract position};
		\node [invisible, below of=pos,yshift=1.3cm] (inv) {};
		
		\draw [arrow] (start) --node[anchor=east] {\large$\bH^{(u_t)}$}(pro1);
		\draw [arrow] (pro1) -- node[anchor=east] {\large$\bmf^{(u_t)}$}(pro2a);
		\draw [arrow] (pro2a) --node[anchor=east] {\large$\hat{\bmp}^{(u_t)}$} (pos);
		\draw [arrow] (pos) --node[anchor=east] {\large$\hat\bmx^{(u_t)}$}(inv) ;
		
		\end{tikzpicture}
	}
} \hspace{0.2cm}
\subfloat[\label{fig:featurelearning}]
{
	\adjustbox{max width=0.28\columnwidth,valign=t}{%
		\begin{tikzpicture}[font=\small,node distance=3cm]
		
		\node (start) [startstop,fill=bl1] {\large{Access point}};
		\node (pro1) [process, below of=start,yshift=1cm,fill=bl5] {\large Feature learning $f_{\boldsymbol\gamma}(\cdot)$};
		\node (pro2a) [process, below of =pro1,yshift=1cm,fill=magenta,fill=bl3] {\large Neural network $g_{\boldsymbol\theta}(\cdot)$};
		\node (pos) [process, below of =pro2a,yshift=1cm,fill=bl4] {\large Extract position};
		\node [invisible, below of=pos,yshift=1.3cm] (inv) {};
		
		\draw [arrow] (start) --node[anchor=east] {\large$\bH^{(u_t)}$}(pro1);
		\draw [arrow] (pro1) -- node[anchor=east] {\large$\bar \bmf^{(u_t)}$}(pro2a);
		\draw [arrow] (pro2a) --node[anchor=east] {\large$\hat{\bmp}^{(u_t)}$} (pos);
		\draw [arrow] (pos) --node[anchor=east] {\large$\hat\bmx^{(u_t)}$}(inv) ;
		
		\end{tikzpicture}
	}
}
\caption{Overview of three different CSI-based positioning pipelines: (a) Estimated CSI is directly fed into a DNN to produce a probability map; (b) estimated CSI is first transformed into CSI features using a pre-designed feature extraction function $f(\cdot)$; (c) proposed architecture that uses a learned feature extraction function $f_{\boldsymbol\gamma}(\cdot)$, which is implemented as a structured DNN.}
\label{fig:systemmodel}
\end{figure}
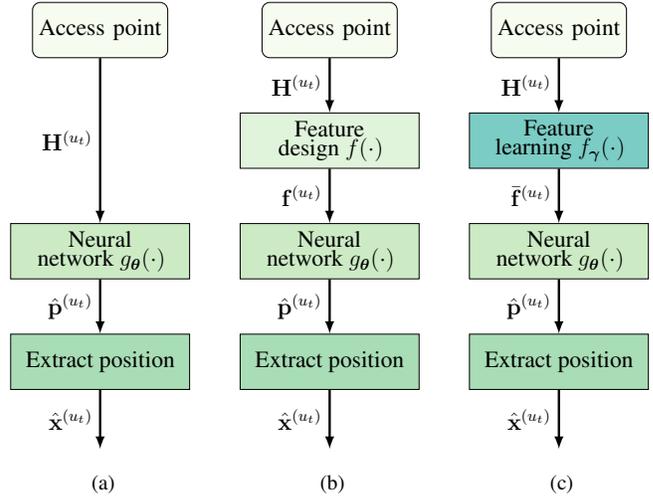

\subsection{CSI-Based Positioning Using DNNs}

Multiple architectures are possible to perform CSI-based positioning with DNNs.  \fref{fig:systemmodel} illustrates three different alternatives, which we detail next.

The first architecture is shown in \fref{fig:rawcsi}. 
The estimated CSI~$\bH^{(u_t)}$ of UE~$u$ at time~$t$ is fed directly to a DNN that computes a probability map $\hat{\bmp}^{(u_t)}\in \reals^K$ according to $\hat \bmp^{(u_t)} = g_{\boldsymbol\theta}(\bH^{(u_t)})$. Here, $K$ is the number of grid points and $g_{\boldsymbol\theta}$ is the function that maps estimated CSI to a probability map~\cite{emrepaper}.
Each grid point $\bmg_k\in\reals^D$, $k=1,\ldots,K$, corresponds to a fixed location in the area of interest, and each entry $\hat{p}^{(u_t)}_k\in[0,1]$, $k=1,\ldots,K$, of the estimated probability map $\hat{\bmp}^{(u_t)}$ indicates the likelihood of the UE $u$ being at the associated grid point at time $t$.
By computing $\hat\bmx^{(u_t)} = \sum_{k=1}^K \bmg_k \hat{p}^{(u_t)}_k$, one then generates an estimate of the UE's position.\footnote{Alternative approaches that directly generate a position estimate within the DNN have been shown in~\cite{emrepaper}  to result in inferior accuracy.}
We emphasize that directly feeding CSI estimates into the DNN does, in general, not perform well (see our results in~\fref{sec:experimentalresults}), mainly because it would require a prohibitive amount of training data to capture a representative set of CSI measurements for all possible system and hardware impairment realizations.

The second architecture is shown in \fref{fig:featuredesign} and overcomes the main drawbacks of the first architecture.
The difference to \fref{fig:rawcsi} is that the CSI estimate $\bH^{(u_t)}\in \complexset^{\MR \times W}$ is first transformed into a CSI feature vector $\bmf^{(u_t)}\in\reals^S$ containing~$S$ features and is computed according to $\bmf^{(u_t)} = f(\bH^{(u_t)})$, where the function $f$ represents the feature extraction stage. 
The feature extraction function $f$ is typically designed by hand (hence dubbed ``feature design'' in \fref{fig:featuredesign}) and tailored to the specifics of the communication system and designed to cope with small-scale fading, antenna orientation, and hardware impairments.
Feature extraction also serves the purpose of preparing the data for subsequent DNN processing~\cite{huawei2020paper}. 
While such architectures are widely used in state-of-the-art CSI-based positioning pipelines~\cite{larsson2015fingerprinting,channelcharting,saidcc,ericpaper,penghzipaper,huawei2020paper,huawei2020channelcharting,emrepaper,foliadis2021csibased}, such hard-coded CSI feature extraction stages are unable to exploit the learning capabilities of DNNs.

The third architecture is shown in \fref{fig:featurelearning}. The key difference to the previous architectures is the fact that we directly \emph{learn} a suitable CSI feature extraction stage.
Specifically, we learn a function $f_{\boldsymbol\gamma}$ (hence dubbed ``feature learning'' in \fref{fig:featurelearning}) with parameters $\boldsymbol\gamma$ that maps CSI estimates to CSI features according to $\bar{\bmf}^{(u_t)} = f_{\boldsymbol\gamma}(\bH^{(u_t)})$. 
By using a \emph{structured} DNN to implement the function  $f_{\boldsymbol\gamma}$, this architecture is able to fully benefit from the power of DNNs, which results in improved position accuracy (see our results in~\fref{sec:experimentalresults}). 

\subsection{Existing CSI Feature Extraction Pipelines}
\label{sec:csipositioning}
As explained above, CSI-based positioning pipelines often rely on hand-crafted CSI features that are extracted from estimated CSI (see \fref{fig:featuredesign}). The CSI features must not only be resilient to small-scale fading, antenna orientation, and hardware impairments, but they should also be different for different transmit positions. 
We now summarize the key steps of the CSI feature extraction pipeline from~\cite{emrepaper}. The main insights will then be used for CSI feature learning in~\fref{sec:featurelearning}.

\subsubsection{Delay-Domain Transform}
\label{sec:beamspace_delaydomain_transform}
Since the propagation channel is described in a compact manner in the time domain, it is often beneficial to convert the frequency-domain CSI into the delay domain according to~\cite{lundpaper,ericpaper,huawei2020paper,emrepaper} 
\begin{align}
\label{eq:delaydomaintransform}
\hat \bH^{(u_t)}=  \bH^{(u_t)} \bF_W^H.
\end{align}
Here, $\bF_W$ is the $W$-dimensional unitary discrete Fourier transform (DFT) matrix\footnote{
Performing an inverse DFT over only the used subcarriers works well in practice and requires lower complexity than first extrapolating the channel coefficients to all subcarriers, e.g., using the method from~\cite{haene2007ofdm}.} and $\hat \bH^{(u_t)}  \in\complexset^{\MR\times W}$ is the delay-domain matrix, which contains the~$C$ taps of the wireless channel between UE $u$ and one AP antenna in each row.

\subsubsection{Autocorrelation}
\label{sec:autocorrelation}
Computing the autocorrelation~\cite{huawei2020paper} or ``instantaneous'' autocorrelation~\cite{emrepaper} has shown to improve resilience of the CSI features to synchronization errors and residual phase errors (caused, e.g., by carrier frequency offset or phase noise). 
Following~\cite{emrepaper}, let $\hat H^{(u_t)}_{m,k}$ be the $k$th delay-domain sample measured at the $m$th receive antenna of UE~$u$ at time $t$. Then, we calculate the 2-dimensional convolution over the delay and antenna domains as 
\begin{align}
\label{eq:autocorrelation_features}
& R^{(u_t)}[\kappa,\tau]  = \sum_{m=1}^\MR \sum_{k=1}^{W}  \hat H^{(u_t)}_{m,k}  \hat H^{(u_t)^*}_{m+\kappa-\MR,k+\tau-W},
\end{align}
where $\kappa=1,2,\ldots,2\MR$ and $\tau=1,2,\ldots,2W$, and the resulting matrix is denoted by~$\bR^{(u_t)}$.

\subsubsection{Real-Valued Decomposition and CSI Feature Normalization}
\label{sec:realvalueddecomposition}
Since some deep learning frameworks are unable to deal with complex numbers, we first vectorize the autocorrelation matrix in \fref{eq:autocorrelation_features} via  $\bmr^{(u_t)} = \mathrm{vec}(\bR^{(u_t)})$, and then convert the $2\MR2W$-dimensional vector into the real domain as follows: 
\begin{align} \label{eq:CSIfeatures}
\hat \bmf^{(u_t)} = \Big[  \Re\{\bmr^{(u_t)}\}^T , \Im\{\bmr^{(u_t)}\}^T \Big]^T.
\end{align} 
In practice, the UE's transmit power and the gain of the low-noise amplifier at the AP are typically set independently. In addition, the path loss can vary vastly for different propagation scenarios. Thus, the received power is, in general, not a reliable quantity for CSI-based indoor positioning applications. 
To this end, we normalize the CSI features in \eqref{eq:CSIfeatures} according to 
$\bmf^{(u_t)} = {\hat \bmf^{(u_t)}}/{\|\hat \bmf^{(u_t)}\|_2}$ prior to feeding them into the DNN~\cite{ericpaper,penghzipaper,olavmultipoint,howardcc,agostini2020cc,emrepaper}. This normalization step also improves convergence of stochastic gradient descent.

\subsection{DNN Structure and Training}
\label{sec:nn}

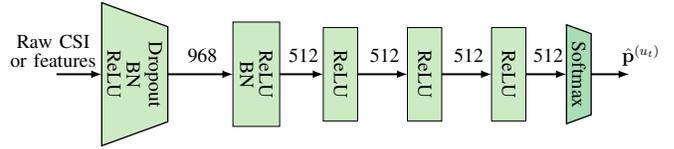
\begin{figure}
\centering
	\adjustbox{max width=\columnwidth,valign=t}{%
		\begin{tikzpicture}[node distance=0cm]
		\node [invisible] (inv) {};
		\node (tr1) [trapezium, trapezium angle=70, draw, thick,minimum width=5mm, minimum height=10mm, outer sep=0pt,text centered, text width=1.8cm,rotate=-90,minimum size=11mm,fill=bl3,above of=inv,yshift=1.8cm] {\large Dropout \\ BN \\ ReLU};
		
		\node (layer1) [process,draw, 
		minimum width=15mm, minimum height=10mm, outer sep=0pt,text centered, text width=2cm,
		xshift=2.6cm,,yshift=-0cm,fill=bl3,above of =tr1,rotate=-90,align=center] {\large  ReLU \\ BN}; 
		
		\node (layer2) [process,draw, 
		minimum width=5mm, minimum height=20mm, outer sep=0pt,text centered, text width=0.5cm,
		xshift=1.8cm,,fill=bl3,right of = layer1] {\large \rotatebox{-90}{ReLU}}; 
		
		\node (layer3) [process,draw, 
		minimum width=5mm, minimum height=20mm, outer sep=0pt,text centered, text width=0.5cm,
		xshift=1.8cm,,fill=bl3,right of = layer2] {\large \rotatebox{-90}{ReLU}}; 
		
		\node (layer4) [process,draw, 
		minimum width=5mm, minimum height=20mm, outer sep=0pt,text centered, text width=0.5cm,
		xshift=1.8cm,,fill=bl3,right of = layer3] {\large \rotatebox{-90}{ReLU}}; 
		
		\node (tr2)[trapezium, trapezium angle=70, draw, thick,minimum width=4mm, minimum height=9mm, outer sep=0pt,text centered, text width=1.5cm,rotate=-90,minimum size=5mm,fill=bl4,above of=layer4, yshift=1.5cm] {\large Softmax};
		
		\node [invisible, right of= tr2,xshift=-1.8cm] (inv2) {};
		
		\draw [arrow] (inv) --node[anchor=east,xshift=0.5cm,yshift=0.5cm]
		{\large \shortstack{Raw CSI \\ or features}}(tr1) ;
		\draw [arrow] (tr1) --node[anchor=east,xshift=0.5cm,yshift=0.4cm] {\large 968}(layer1) ;
		\draw [arrow] (layer1) --node[anchor=east,xshift=0.5cm,yshift=0.4cm] {\large 512 }(layer2) ;
		\draw [arrow] (layer2) --node[anchor=east,xshift=0.5cm,yshift=0.4cm] {\large 512}(layer3) ;
		\draw [arrow] (layer3) --node[anchor=east,xshift=0.5cm,yshift=0.4cm] {\large 512}(layer4) ;
		\draw [arrow] (layer4) --node[anchor=east,xshift=0.5cm,yshift=0.4cm] {\large 512}(tr2) ;
		
		\draw [arrow] (tr2) --node[anchor=east,xshift=1.2cm,yshift=0.4cm] {\large $\hat{\bmp}^{(u_t)}$}(inv2) ;
		
		\end{tikzpicture}
	}
\vspace{-0.1cm}	
\caption{DNN topology used to generate probability maps from raw estimated CSI $\bH^{(u_t)}$, designed CSI features $\bmf^{(u_t)}$, or learned CSI features $\bar \bmf^{(u_t)}$.}
\label{fig:nnmodel}
\end{figure}

For all of the positioning pipelines shown in \fref{fig:systemmodel}, we use the same DNN structure from~\cite{emrepaper}, as illustrated in \fref{fig:nnmodel}.
For the architecture depicted in \fref{fig:rawcsi}, the positioning DNN takes in raw CSI measurements. For the architecture  depicted in \fref{fig:featuredesign}, the positioning DNN takes in hand-designed features. For the architecture depicted in \fref{fig:featurelearning}, the positioning DNN additionally includes trainable input layers that implement the feature learning function $f_\gamma$ which processes raw CSI measurements; the rest of the positioning DNN is the same as for the two other architectures. 
The positioning DNN $g_{\boldsymbol\theta}$, where the vector~$\boldsymbol\theta$ contains all weights and biases, consists of six hidden layers and the numbers in \fref{fig:nnmodel} indicate the number of activations per layer. 
All layers use rectified linear unit (ReLU) activation functions, except the last layer is using  a softmax activation to generate the estimated probability map~$\hat \bmp^{(u_t)}$ for the $u$th UE at time~$t$. 
The first layer uses batch normalization (BN) and dropout; the second layer only uses BN.

We assume that all of the DNNs are trained off-line from a dataset containing CSI-location pairs. While self-supervised methods, e.g., channel charting~\cite{channelcharting}, could avoid the acquisition of large data sets, we leave an investigation of such methods for future work. 
During training, we initialize the weights using Glorot~\cite{glorot} and use the binary cross entropy (BCE) as the loss. 
To generate the training set, we compute a reference probability map~$\bmp^{(u_t)}$ using the approach described in~\cite{emrepaper} for each reference position~$\bmx^{(u_t)}$. Concretely, we identify the nearest four grid points to the reference position and assign the associated four probability values so that the expected position corresponds to the reference position.

\section{CSI Feature Learning}
\label{sec:featurelearning}

We now propose our approach to learn the CSI feature extraction stage as shown in \fref{fig:featurelearning} instead of working with hard-coded CSI features. 
We note that learning an unstructured DNN directly from the raw CSI features would suffer from the same limitations as the architecture in \fref{fig:rawcsi}.
Thus, we use a simple but effective alternative: We first build a structured DNN that models the key steps of the feature extraction pipeline detailed in \fref{sec:csipositioning}. We then learn the key parameters of this structured DNN together with the DNN that generates the probability maps. 
These steps are detailed next. 

\subsection{Delay-Domain Transform}
We apply the transform in \fref{eq:delaydomaintransform} with a trainable weight matrix $\bW_1\in\opC^{W\times W}$ instead of the inverse DFT (IDFT) matrix $\bF_W^H$. 
To this end, we initialize the trainable weight matrix with~$\bF_W^H$.
We then compute $\hat \bH^{(u_t)} = \bH^{(u_t)}\bW_1$ using the real-valued decomposition.
We note that after learning the weight matrix~$\bW_1$, the transformed matrix~$\hat \bH^{(u_t)}$ contains CSI features that are not necessarily in the delay domain anymore. 
\subsection{DFT-Based Autocorrelation}
In order to  calculate the instantaneous 2D autocorrelation in \eqref{eq:autocorrelation_features}, we use the Wiener–Khinchin theorem. Specifically, we calculate the autocorrelation matrix with the aid of DFT and IDFT matrices as follows:
\begin{align}
\label{eq:wienerkhinchin}
\hat \bR^{(u_t)}= \bF_{2\MR}^H& |\bF_{2\MR} \bH_z^{(u_t)} \bF_{2W}|^{\circ2}\,\bF_{2W}^H.
\end{align}
Here, $\hat \bR^{(u_t)} \in \complexset^{2\MR\times 2W}$ is the autocorrelation of the zero-padded delay-domain CSI matrix $\bH_z^{(u_t)}\in \complexset^{2\MR\times 2W}$, $\bF_{2\MR}$ and $\bF_{2W}$ are the $2\MR$-dimensional and $2W$-dimensional DFT matrices respectively.
Zero padding is achieved by appending zeros to the columns and rows of $ \hat \bH^{(u_t)}$ to get $\bH_z^{(u_t)}$, which doubles the number of rows and columns.

Our main idea for this stage is to calculate \fref{eq:wienerkhinchin}, where we replace 
the DFT matrix $\bF_{2\MR}$ by a trainable weight matrix $\bW_2\in\complexset^{2\MR\times2\MR}$, which is initialized by $\bF_{2\MR}$, and $\bF_{2\MR}^H$ by $\bW_2^H$.
Analogously, we replace the DFT matrix $\bF_{2W}$ by a trainable weight matrix $\bW_3\in\complexset^{2W\times2W}$, which is initialized by $\bF_{2W}$, and $\bF_{2W}^H$ by $\bW_3^H$.
We then compute
\begin{align}
\label{eq:autocorr} 
\hat \bR^{(u_t)}= \bW_2^H& |\bW_2 \bH_z^{(u_t)} \bW_3|^{\circ2}\,\bW_3^H.
\end{align}
We note that after learning the two weight matrices $\bW_2$ and~$\bW_3$, the matrix $\hat \bR^{(u_t)}$ in  \fref{eq:autocorr} does not necessarily correspond to the autocorrelation anymore.
Once again, we carry out the above steps using the real-valued decomposition.

\subsection{CSI Feature Normalization}
Finally, we vectorize the matrix $\hat \bR^{(u_t)}$ from~\fref{eq:autocorr} via $\hat\bmr^{(u_t)} = \mathrm{vec}(\hat\bR^{(u_t)})$ and separate the real and imaginary parts as in~\fref{eq:CSIfeatures}, resulting in the vector $\hat\bmf^{(u_t)}$.
As a last step, we normalize the result as $\bar\bmf^{(u_t)} = {\hat\bmf^{(u_t)}}/{\|\hat\bmf^{(u_t)}\|_2}$, which is the CSI feature vector that is fed in the positioning DNN (cf.~\fref{fig:featurelearning}).

\begin{figure*}[tp]
	\vspace{-0.7cm}	
	\centering
	\subfloat[\label{fig:featurefusion}]
	{
		\adjustbox{width=0.22\textwidth,valign=t}{%
			\begin{tikzpicture}[font=\small,node distance=3cm]
			
			\node [rectangle, node distance=3cm] (start) {\large$\bH^{(u_{t-\delta})}$};
			\node (pro1) [process, below of=start,yshift=1.5cm,fill=bl5,minimum height=0.5cm] {\large 
				$f_{\boldsymbol\gamma}(\cdot)$};

			\node (rnn) [process, below of =pro1,yshift=1cm,fill=magenta,fill=bl3,text width=8cm,xshift=2.5cm] {\large Feature fusion};
		
			\node [invisible, below of=pos,yshift=1.3cm] (inv) {};

			\draw [arrow] (start) --node[anchor=east] {}(pro1);
			\draw [arrow] (pro1) -- node[anchor=east] {\large$\bar \bmf^{(u_{t-\delta})}$}([xshift=-2.5cm]rnn.north);

			\node [rectangle, node distance=5cm,right of= start] (start1) {\large$\bH^{(u_t)}$};

			\node (pro11) [process, below of=start1,yshift=1.5cm,fill=bl5,minimum height=0.5cm] {\large 
				 $f_{\boldsymbol\gamma}(\cdot)$};
			
			\node (pro2a1) [process, below of =rnn,yshift=1.5cm,xshift=2.5cm,fill=magenta,fill=bl3,,minimum height=0.5cm] {\large 
				$g_{\boldsymbol\theta}(\cdot)$};
			
			\node (pos1) [process, below of =pro2a1,yshift=-1.35cm,fill=bl4] {\large Extract position};
			\node [invisible, below of=pos1,yshift=1.3cm] (inv1) {};
			
			\path (start) -- node[auto=false]{\ldots} (start1);
			\path (pro1) -- node[auto=false]{\ldots} (pro11);
			
			\draw [arrow] (start1) --node[anchor=east] {}(pro11);
			\draw [arrow] (pro11) -- node[anchor=east] {\large$\bar \bmf^{(u_t)}$}([xshift=2.5cm]rnn.north);
			\draw [arrow] ([xshift=2.5cm]rnn.south) -- node[anchor=east] {\large $\underline\bmf^{(u_t)}$}(pro2a1);
			\draw [arrow] (pro2a1) --node[anchor=east] {\large$\hat \bmp^{(u_t)}$} (pos1);
			\draw [arrow] (pos1) --node[anchor=east] {\large$\hat\bmx^{(u_t)}$}(inv1) ;
			
			\end{tikzpicture}
		}
	}
	~~~~
	\subfloat[\label{fig:probabilitymapfusion}]
	{
		\adjustbox{width=0.22\textwidth,valign=t}{%
			\begin{tikzpicture}[font=\small,node distance=3cm]

			\node [rectangle, node distance=3cm] (start) {\large$\bH^{(u_{t-\tau})}$};

			\node (pro1) [process, below of=start,yshift=1.5cm,fill=bl5,minimum height=0.5cm] {\large 
				 $f_{\boldsymbol\gamma}(\cdot)$};

			\node (pro2a) [process, below of =pro1,yshift=-0.5cm,fill=magenta,fill=bl3,minimum height=0.5cm] {\large
				 $g_{\boldsymbol\theta}(\cdot)$};
			
			\node (rnn) [process, below of =pro2a,yshift=0.95cm,fill=magenta,fill=bl3,text width=8cm,xshift=2.5cm] {\large Probability map fusion};

			\node [invisible, below of=pos,yshift=1.3cm] (inv) {};

			\draw [arrow] (start) --node[anchor=east] {}(pro1);
			\draw [arrow] (pro1) -- node[anchor=east] {\large$\bar \bmf^{(u_{t-\tau})}$}(pro2a);
			\draw [arrow] (pro2a) --node[anchor=east] {\large$\hat \bmp^{(u_{t-\tau})}$} ([xshift=-2.5cm]rnn.north);

			\node [rectangle, node distance=5cm,right of= start] (start1) {\large$\bH^{(u_t)}$};
			
			\node (pro11) [process, below of=start1,yshift=1.5cm,fill=bl5,minimum height=0.5cm] {\large 
				 $f_{\boldsymbol\gamma}(\cdot)$};

			\node (pro2a1) [process, below of =pro11,yshift=-0.5cm,fill=magenta,fill=bl3,minimum height=0.5cm] {\large 
				$g_{\boldsymbol\theta}(\cdot)$};
			
			\node (pos1) [process, below of =rnn,yshift=0.7cm, xshift=2.5cm,fill=bl4] {\large Extract position};
			\node [invisible, below of=pos1,yshift=1.3cm] (inv1) {};

			\path (start) -- node[auto=false]{\ldots} (start1);
			\path (pro1) -- node[auto=false]{\ldots} (pro11);
			\path (pro2a) -- node[auto=false]{\ldots} (pro2a1);
			
			\draw [arrow] (start1) --node[anchor=east] {}(pro11);
			\draw [arrow] (pro11) -- node[anchor=east] {\large$\bar \bmf^{(u_t)}$}(pro2a1);
			\draw [arrow] ([xshift=2.5cm]rnn.south) -- node[anchor=east] {\large$\underline \bmp^{(u_t)}$}(pos1);
			\draw [arrow] (pro2a1) --node[anchor=east] {\large$ \hat \bmp^{(u_t)}$} ([xshift=2.5cm]rnn.north);
			\draw [arrow] (pos1) --node[anchor=east] {\large$ \hat \bmx^{(u_t)}$}(inv1) ;
			
			\end{tikzpicture}
		}
	}
	~~~~
	\subfloat[\label{fig:allfusion}]
	{
		\adjustbox{ width=0.485\textwidth,valign=t}{%
			\begin{tikzpicture}[font=\small,node distance=3cm]

			\node [rectangle, node distance=3cm] (start) {\large$\bH^{(u_{t-\tau-\delta})}$};

			\node (pro1) [process, below of=start,yshift=1.5cm,fill=bl5,minimum height=0.5cm] {\large 
				 $f_{\boldsymbol\gamma}(\cdot)$};

			\node (rnn) [process, below of =pro1,yshift=1cm,fill=magenta,fill=bl3,text width=8cm,xshift=2.5cm] {\large Feature fusion};
			
			\node [invisible, below of=pos,yshift=1.3cm] (inv) {};

			\draw [arrow] (start) --node[anchor=east] {}(pro1);
			\draw [arrow] (pro1) -- node[anchor=east] {\large$\bar \bmf^{(u_{t-\tau-\delta})}$}([xshift=-2.5cm]rnn.north);

			\node [rectangle, node distance=5cm,right of= start] (start1) {\large$\bH^{(u_{t-\tau})}$};
			
			\node (pro11) [process, below of=start1,yshift=1.5cm,fill=bl5,minimum height=0.5cm] {\large 
				 $f_{\boldsymbol\gamma}(\cdot)$};
			\node (pro2a1) [process, below of =rnn,yshift=1.5cm,xshift=2.5cm,fill=magenta,fill=bl3,minimum height=0.5cm] {\large 
				$g_{\boldsymbol\theta}(\cdot)$};

			\node [invisible, below of=pos1,yshift=1.3cm] (inv1) {};
			
			\path (start) -- node[auto=false]{\ldots} (start1);
			\path (pro1) -- node[auto=false]{\ldots} (pro11);

			\draw [arrow] (start1) --node[anchor=east] {}(pro11);
			\draw [arrow] (pro11) -- node[anchor=east] {\large$\bar \bmf^{(u_{t-\tau})}$}([xshift=2.5cm]rnn.north);
			\draw [arrow] ([xshift=2.5cm]rnn.south) -- node[anchor=east] {\large $\underline\bmf^{(u_{t-\tau})}$}(pro2a1);

			\node [rectangle, node distance=5cm,right of=start1] (start21) {\large$\bH^{(u_{t-\delta})}$};

			\node (pro21) [process, below of=start21,yshift=1.5cm,fill=bl5,minimum height=0.5cm] {\large 
				 $f_{\boldsymbol\gamma}(\cdot)$};
			
			\node (rnn21) [process, below of =pro21,yshift=1.0cm,fill=magenta,fill=bl3,text width=8cm,xshift=2.5cm] {\large Feature fusion};

			\node (pro2a22) [process, below of =rnn21,yshift=1.5cm,xshift=2.5cm,fill=magenta,fill=bl3,minimum height=0.5cm] {\large 
				 $g_{\boldsymbol\theta}(\cdot)$};
			
			\node [invisible, below of=pro2a22,yshift=1.3cm] (inv21) {};

			\draw [arrow] (start21) --node[anchor=east] {}(pro21);
			\draw [arrow] (pro21) -- node[anchor=east] {\large$\bar \bmf^{(u_{t-\delta})}$}([xshift=-2.5cm]rnn21.north);

			\node [rectangle, node distance=5cm,right of= start21] (start22) {\large$\bH^{(u_t)}$};
			\node (pro22) [process, below of=start22,yshift=1.5cm,fill=bl5,minimum height=0.5cm] {\large 
				 $f_\gamma(\cdot)$};

			\node (rnnoutput) [process, below of =pro2a22,yshift=1cm,fill=magenta,fill=bl3,text width=18cm,xshift=-7.5cm] {\large Probability map fusion};
			
			\draw [arrow] (pro2a1) -- node[anchor=east] {\large$ \hat \bmp^{(u_{t-\tau})}$}([xshift=-2.5cm]rnnoutput.north);
			
			\draw [arrow] (pro2a22.south) -- node[anchor=east] {\large$ \hat \bmp^{(u_t)}$}([xshift=7.5cm]rnnoutput.north);

			\node (posextract) [process, below of =rnnoutput,yshift=0.7cm,fill=bl4,xshift=7.5cm] {\large Extract position};
			
			\node [invisible, below of=posextract,yshift=1.3cm] (inv22) {};
			
			\path (start21) -- node[auto=false]{\ldots} (start22);
			\path (pro21) -- node[auto=false]{\ldots} (pro22);
			
			\draw [arrow] (start22) --node[anchor=east] {}(pro22);
			\draw [arrow] (pro22) -- node[anchor=east] {\large$\bar \bmf^{(u_t)}$}([xshift=2.5cm]rnn21.north);
			\draw [arrow] ([xshift=2.5cm]rnn21.south) -- node[anchor=east] {\large $\underline\bmf^{(u_{t})}$}(pro2a22);
			\draw [arrow] ([xshift=7.5cm]rnnoutput.south) --node[anchor=east] {\large$\underline \bmp^{(u_t)}$} (posextract);
			\draw [arrow] (posextract) --node[anchor=east] {\large$\hat \bmx^{(u_t)}$}(inv22) ;

			\path (pro2a1) -- node[auto=false]{\Large\ldots} (pro2a22);
			\path (rnn) -- node[auto=false]{\Large\ldots} (rnn21);
			\path (pro11) -- node[auto=false]{\Large \ldots} (pro21);

			\end{tikzpicture}
		}
	}	
\caption{CSI feature fusion and probability map fusion: (a) CSI feature fusion in which an RNN combines a time-series of CSI estimates to produce a CSI feature; (b) probability map fusion in which an RNN or Gaussian conflation combines a time-series of probability maps; (c) CSI feature fusion and probability map fusion that combines both methods to compute a combined probability map.}
\label{fig:fusion}
\end{figure*}
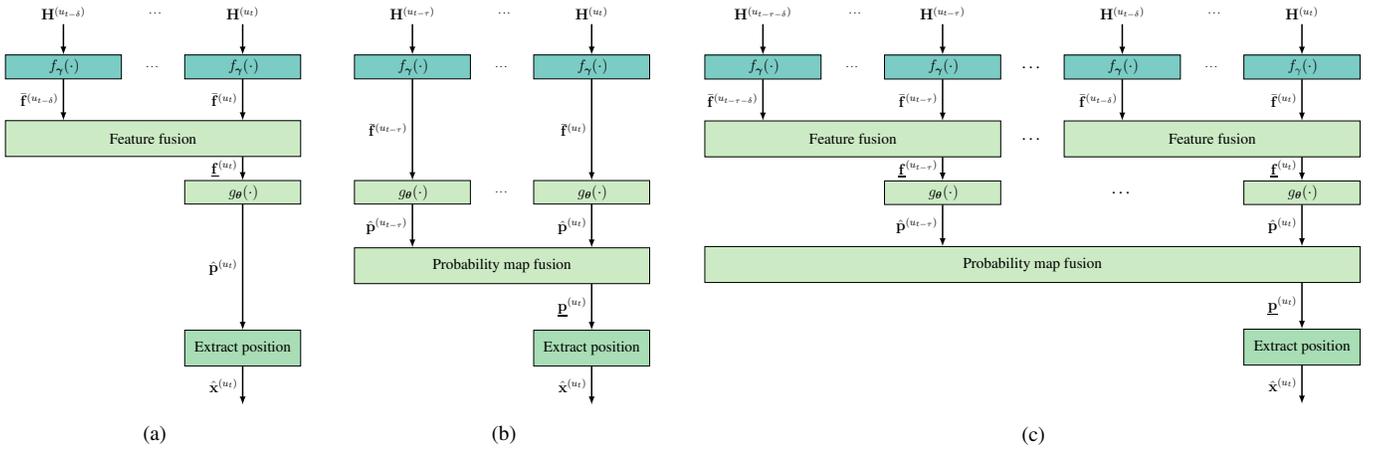

\section{Time-Fusion of CSI-Features \\ and Probability Maps}
\label{sec:timefusion}
We now propose three methods that take into account the facts that (i) CSI estimates are generated at fast rates and (ii) multiple consecutive CSI features and/or probability maps can be fused to improve positioning accuracy of moving UEs.

\subsection{Fusion of CSI Features}
\label{sec:featurefusion}

In practical situations, CSI estimates are generated as a time series and consecutive CSI features will exhibit some degree of similarity due to the relatively slow motion of the UEs with respect to the rate of CSI acquisition. 
Our idea is illustrated in \fref{fig:featurefusion} and fuses $\delta+1$ consecutive CSI estimates $\{\bH^{(u_{n})}\}_{n=t-\delta}^{t}$ using a recurrent neural network (RNN).
For each time step, we first apply our trainable CSI feature extraction layer to each CSI estimate, followed by a CSI feature fusion layer that is implemented using $\delta+1$ gated-recurrent-units (GRUs)~\cite{gru} with linear activation functions. 
The GRUs produce a single CSI feature vector $\underline{\bmf}^{(u_t)}$ that is passed to the neural network $g_{\boldsymbol\theta}$ to produce a probability map.

\subsection{Probability Map Fusion}
\label{sec:probabilitytimefusion}
An alternative to fusing time series of CSI features is to fuse $\tau+1$ consecutive  probability maps $\{ \hat \bmp^{u_{n}}\}_{n=t-\tau}^{t}$. 
The basic idea is illustrated in \fref{fig:probabilitymapfusion} and we propose two different fusion methods. 
The first fusion approach uses a simple four hidden-layer RNN with an initial layer consisting of $\tau+1$ GRUs~\cite{gru} with $K$ linear activation functions, which takes in the time series of probability maps and computes a fused probability map $\underline{\bmp}^{u_{t}}$.
The hidden layers have $K$ ReLU activations, and the output layer has a softmax activation with~$K$ outputs.
The second fusion approach uses Gaussian conflation~\cite{hill2008conflations}, which has been used in~\cite{emrepaper} to fuse probability maps from multiple APs and multiple transmit antennas to improve positioning accuracy. 
Here, we first compute the location estimates $\{\hat\bmx^{(u_n)}\}_{n=t-\tau}^{t}$ and associated covariance matrices $ \bK^{(u_n)}= \sum_{k=1}^K  p^{(u_n)}_k (\bmg_k - \hat\bmx^{(u_n)}) (\bmg_k - \hat\bmx^{(u_n)})^T$ for $n=t-\tau,\ldots, t$ from the $\tau+1$  probability maps $\{ \hat \bmp^{u_{n}}\}_{n=t-\tau}^{t}$. We then perform Gaussian conflation~\cite{emrepaper}
\begin{align}\label{eq:Gaussianconflation}
\hat x^{(u_t)}_d = \frac{\sum_{n=t-\tau}^{t} K^{(u_n)^{-1}}_{d,d} \hat x_d^{(u_n)}}{\sum_{n=t-\tau}^{t} K^{(u_n)^{-1}}_{d,d}  },
\end{align}
where $d=1,\ldots,D$ with $D=2$.
The intuition behind \fref{eq:Gaussianconflation} is that location estimates with large variance (which are less reliable) are deweighted in the averaging process.

\subsection{CSI Feature and Probability Map Fusion}
\label{sec:allfusion}
In order to further improve the positioning accuracy, we can combine CSI feature fusion with probability map fusion. The idea is illustrated in \fref{fig:allfusion} and takes in  $\tau+1$ separate groups of raw CSI estimates. Each group of $\delta+1$ consecutive raw CSI estimates are first passed through our learned CSI feature extraction pipeline and feature fusion RNN. The features are then passed through DNNs that produce $\tau+1$ probability maps, which are fused using an RNN or Gaussian conflation in order to produce a fused probability map.
Finally, this probability map is used to compute the position estimate.

\section{Results}
\label{sec:results}

We now provide experimental results for the CSI feature learning and time fusion methods proposed in  \fref{sec:featurelearning} and \fref{sec:timefusion}, respectively. We start by describing the measurement setup and then show our results.

\subsection{Measurement Setup}

For our experiments, we use two datasets from~\cite{emrepaper} containing real-world CSI measurements recorded in a LoS lab scenario (called \textit{LoS Lab v1} with 154k training samples and 3800 test samples) and a non-LoS home scenario (called \textit{NLoS Home} with 96k training samples and 2200 test samples).
In addition, we also acquired a new dataset (called \textit{LoS Lab Data v2} with 202k training samples and 50k test samples), which has been recorded in a lab space at Cornell University using the same measurement setup described in~\cite{emrepaper}.
The dataset is generated with a robot that follows a random path in a 3.2$\times$4.2\,m$^\text{2}$ area under LoS conditions, where we first record the training set and then a test set that have different UE locations.
This new dataset is more challenging than the two other datasets from~\cite{emrepaper} as the test set contains many locations that were not previously fingerprinted in the training set. 
We obtain CSI measurements for $W=234$ subcarriers and at $\MR=4$ receive antennas, and we use a VICON~\cite{vicon} precision positioning system to collect the ground-truth location information.
To construct the probability maps, we use a $22\times22$ uniformly-spaced grid which corresponds to $K=484$ grid-point probabilities. 
In what follows, we provide the mean distance error (MDE) in meters evaluated on the test sets.

\begin{table}[t]
\vspace{-0.2cm}
	\centering
	\caption{Impact of CSI feature learning on MDE.}
	\vspace{-0.1cm}
	\label{tbl:featurelearn}
	\renewcommand{\arraystretch}{1.06}
	\resizebox{0.65\columnwidth}{!}{
	\begin{tabular}{@{}lccc@{}}
		\toprule
		\textbf{Features}             & \textbf{ Raw CSI} & \textbf{Designed} & \textbf{Learned } \\
		Figure & \ref{fig:rawcsi} & \ref{fig:featuredesign} & \ref{fig:featurelearning}\\
		 \midrule
		LoS Lab v1    & 0.36    & 0.05             & \textbf{0.04}                    \\
		LoS Lab v2 & 0.50     & 0.30              & \textbf{0.25}                    \\
		NLoS Home      & 0.52    & 0.28           & \textbf{0.19}  \\
		\bottomrule                  
	\end{tabular}} \\[0.4cm]
	\caption{Impact of CSI feature time-fusion on MDE.}
	\vspace{-0.1cm}
	\label{tbl:featurefusion}
	\renewcommand{\arraystretch}{1.06}
	\resizebox{0.8\columnwidth}{!}{
	\begin{tabular}{@{}lccc@{}}
		\toprule
		\textbf{Features} & \textbf{Non-fusion, learned} & \textbf{Designed} & \textbf{Learned} \\
		Figure & \ref{fig:featurelearning} & \ref{fig:featurefusion} and \ref{fig:featuredesign} & \ref{fig:featurefusion} and \ref{fig:featurelearning}  \\
		 \midrule
		LoS Lab v1       &  0.04   & 0.04        & \textbf{0.03}                                                    \\
		LoS Lab v2       &  0.25    & \textbf{0.24}                                                       & \textbf{0.24}                                                     \\
		NLoS Home        &  0.19    & 0.21                                                       & \textbf{0.17}   \\ \bottomrule                                                 
	\end{tabular}} \\[0.4cm]
	\caption{Impact of probability map fusion on MDE.}
	\vspace{-0.1cm}
	\label{tbl:outputrnn}
	\renewcommand{\arraystretch}{1.06}
	\resizebox{\columnwidth}{!}{
	\begin{tabular}{@{}lccccc@{}}
		\toprule
		~ & \textbf{Non-fusion}&\multicolumn{2}{c}{\textbf{RNN} }& \multicolumn{2}{c}{\textbf{Gaussian conflation}} \\
		\textbf{Features}& Learned &Designed  & Learned 
		&Designed  & Learned   
		\\ 
		Figure  & \ref{fig:systemmodel}&  \ref{fig:probabilitymapfusion} and \ref{fig:featuredesign}  &  \ref{fig:probabilitymapfusion} and \ref{fig:featurelearning} &   \ref{fig:probabilitymapfusion} and \ref{fig:featuredesign}&  \ref{fig:probabilitymapfusion} and \ref{fig:featurelearning} \\
		\midrule
		LoS Lab v1    &0.04              & \textbf{0.03   }                              & \textbf{0.03}             & \textbf{0.03}
		&\textbf{0.03}                \\
		LoS Lab v2       & 0.25            & 0.23                                  & \textbf{0.19}            & 0.23
		& 0.20                     \\
		NLoS Home        &0.19            & 0.22                                 & \textbf{0.16}           &0.23 & 0.17             \\ \bottomrule       
	\end{tabular}} \\[0.4cm]
	\caption{Impact of CSI feature and probability map fusion on MDE.}
	\vspace{-0.1cm}
	\label{tbl:rnninputoutput}
	\renewcommand{\arraystretch}{1.06}
	\resizebox{\columnwidth}{!}{
	\begin{tabular}{@{}lcccccc@{}}
		\toprule
		~ & \textbf{Non-fusion}& \multicolumn{2}{c}{\textbf{RNN} }& \multicolumn{2}{c}{\textbf{Gaussian conflation}} \\
		\textbf{Features}& Learned & Designed & Learned
		&Designed  & Learned   \\
		Figure  & \ref{fig:featurelearning} & \ref{fig:allfusion} and \ref{fig:featuredesign}  & \ref{fig:allfusion} and \ref{fig:featurelearning}  & \ref{fig:allfusion} and \ref{fig:featuredesign}& \ref{fig:allfusion} and \ref{fig:featurelearning} \\
		\midrule
		LoS Lab v1       &0.04                  & 0.04                                                    & \textbf{0.02}     &0.04
		&\textbf{0.02}                                            \\
		LoS Lab v2   &0.25                       & 0.22                                                      & \textbf{0.21}                                     &\textbf{0.21}
		&	\textbf{0.21}               \\
		NLoS Home     &0.19                          & 0.20                                                   & \textbf{0.15}     &0.19
		&
		\textbf{0.15}                \\ \bottomrule                                
	\end{tabular}}
\end{table}

\subsection{Experimental Results}
\label{sec:experimentalresults}

We now show our experimental results for the proposed CSI feature learning and time fusion methods.

\subsubsection{Impact of CSI Feature Learning}
%
Table \ref{tbl:featurelearn} shows MDE results for the following methods: (i)  \fref{fig:rawcsi}, in which the DNN~$g_{\boldsymbol\theta}$ takes in \textit{raw} CSI measurements and generates probability maps as described in \fref{sec:featurelearning}; (ii) \fref{fig:featuredesign}, in which the DNN~$g_{\boldsymbol\theta}$ uses \emph{designed} CSI features as described in \fref{sec:csipositioning}; and (iii) \fref{fig:featurelearning}, in which the neural network~$g_{\boldsymbol\theta}$ takes in the output of the proposed CSI \emph{feature learning} approach as described in \fref{sec:featurelearning}.
For the \textit{LoS Lab v1} and \textit{NLoS Home} scenarios, the MDE for the designed CSI feature pipeline matches those reported in~\cite{emrepaper}. 
As mentioned earlier, the use of raw CSI estimates does not perform well.
The proposed CSI feature learning approach, however, consistently outperforms the two other methods in terms of the MDE.

\subsubsection{Impact of CSI Feature Time-Fusion}
%
Table \ref{tbl:featurefusion} shows MDE results for CSI feature time fusion  as  detailed in \fref{sec:featurefusion}. Here, the RNN in \fref{fig:featurefusion} takes in $\delta+1=3$ subsequent CSI features to generate a combined CSI feature that is used to generate probability maps (we have observed that $\delta+1=3$ resulted in the lowest MDE).
Compared to the non-fusion case, we observe an additional MDE decrease for all datasets. 

\subsubsection{Impact of Probability Map Time-Fusion}
Table \ref{tbl:outputrnn} shows MDE results for probability map time fusion as detailed in \fref{sec:probabilitytimefusion}. Here, an RNN or Gaussian conflation is used to combine  $\tau+1=3$ subsequent probability maps to generate a combined probability map (we have observed that $\tau+1=3$ resulted in the lowest MDE).
Compared to the non-fusion case, we observe an additional MDE decrease for all datasets, where the MDE is slightly lower for the RNN-based approach (compared to Gaussian conflation) for the \textit{LoS Lab v2} and \textit{NLoS Home} datasets.

\subsubsection{Impact of CSI Feature and Probability Map Time-Fusion}
%
Table \ref{tbl:rnninputoutput} shows MDE results for the combination of CSI feature and probability map fusion as detailed in \fref{sec:allfusion}. Here, we take groups of $\tau+1=3$ probability maps, each of which are generated by fusing groups of $\delta+1=3$ CSI features, and outputs a combined probability map as shown in \fref{fig:allfusion}.
Compared to CSI feature learning alone (no time fusion) and hand-crafted CSI features, we observe a $2\times$ and a $2.5\times$ MDE reduction, respectively. 
Fusing both the CSI features and the probability maps results in the lowest MDE. 

\section{Conclusions}
We have proposed a DNN-based positioning pipeline that takes in raw CSI estimates and learns CSI features using a structured positioning DNN. 
We have also proposed two time-fusion methods, which combine a time series of CSI features and probability maps in order to further improve positioning accuracy.
Our experiments with real-world indoor CSI measurements show a reduction in mean distance error of up to $2.5\times$ compared to state-of-the-art methods that build on hand-crafted CSI features. 
Our next step is to apply our methods to self-supervised channel charting~\cite{channelcharting} and semi-supervised positioning methods~\cite{ericpaper,penghzipaper} in order to improve their localization capabilities. 

	\balance
	

	\balance
	
\end{document}